\documentclass[aps,prd,preprint,superscriptaddress,showpacs]{revtex4-1}

\usepackage{amsmath, amsfonts, amsthm, amssymb, mathrsfs, enumerate, graphicx}

\newcommand{\sech}{\mathrm{sech} \,}
\newcommand{\R}{{\mathbb R}}

\begin{document}


\title{Constructing vacuum spacetimes by generating manifolds of revolution around a curve}


\author{Vee-Liem Saw}
\email[]{VeeLiem@maths.otago.ac.nz}
\affiliation{Department of Mathematics and Statistics, University of Otago, Dunedin 9016, New Zealand}


\date{\today}

\begin{abstract}
We develop a general perturbative analysis on vacuum spacetimes which can be constructed by generating manifolds of revolution around a curve, and apply it to the Schwarzschild metric. The following different perturbations are carried out separately: 1) Non-rotating 2-spheres are added along a plane curve slightly deviated from the ``Schwarzschild line''; 2) General non-rotating topological 2-spheres are added along the ``Schwarzschild line'' 3) Slow-rotating 2-spheres are added along the ``Schwarzschild line''. For (1), we obtain the first order vacuum solution and show that no higher order solution exists. This linearised vacuum solution turns out however to be just a gauge transformation of the Schwarzschild metric. For (2), we solve the general linearised vacuum equations under several special cases. In particular, there exist linearised vacuum solutions with signature-changing metrics that contain closed timelike curves (though these do not correspond to adding topological 2-spheres). For (3), we find that the first order vacuum solution is equivalent to the slowly rotating Kerr metric. This is hence a much simpler and geometrically insightful derivation as compared to the gravitomagnetic one, where this rotating-shells construction is a direct manifestation of the frame-dragging phenomenon. We also show that the full Kerr however, cannot be obtained via adding rotating ellipsoids.
\end{abstract}

\pacs{04.20.Cv, 04.25.Nx}

\maketitle


\sloppy

\section{Introduction}

The general method of constructing spacetime by generating manifolds of revolution around a given curve was originally devised to build static curved traversable wormholes \cite{Vee2012, Vee2013}\footnote{This method of generating manifolds of revolution around a given curve originated from the idea of helicalising an arbitrary smooth curve, i.e. to replace a given curve by a curve which winds around it \cite{Vee2013b}. By introducing a second parameter, the result is a surface of revolution around the given curve. A third parameter yields a 3-manifold of revolution, $\vec{\sigma}(u,v,w)$ in Eq. (\ref{method}).} which do not rely on the assumption of spherical symmetry. A key result found using this method was that although exotic matter is required to support a traversable wormhole \cite{Kip, Vis2}, it is possible to engineer the shape and curvature of curved ones so that safe geodesics through them exist. This allows travellers to traverse on a freely-falling trajectory locally supported by ordinary matter, avoiding the need for direct contact with the exotic matter.

Here is the method. Take a given smooth curve $\vec{\psi}(v)$ embedded into $\R^4$. The 3-manifold of revolution formed by adding 2-spheres along $\vec{\psi}$ is given by:
\begin{eqnarray}\label{method}
\vec{\sigma}(u,v,w)=\vec{\psi}(v)+Z(v)\sin{u}\cos{w}\ \hat{n}_1(v)+Z(v)\sin{u}\sin{w}\ \hat{n}_2(v)+Z(v)\cos{u}\ \hat{n}_3(v),
\end{eqnarray}
where $Z$ is the radial function determining the variation of the 2-spheres' radii along $\vec{\psi}$ and $\hat{n}_1,\hat{n}_2,\hat{n}_3$ are three orthonormal vectors, canonically taken to be perpendicular to $\vec{\psi}$ (unless simpler equations can otherwise be produced). So $\vec{\sigma}$ is a map from the 3-manifold of revolution into $\R^4$. The metric of the 3-manifold can be calculated by pulling back the standard Euclidean metric of $\R^4$ via $\vec{\sigma}$ (see for instance appendix A of Ref. \cite{Carroll}) \footnote{Alternatively, this can also be viewed as embedding the 3-manifold of revolution into flat 4-d Euclidean space $ds^2=dx_1^2+dx_2^2+dx_3^2+dx_4^2$, with $\vec{\sigma}=(x_1,x_2,x_3,x_4)$.}, and subsequently extended to become a (3+1)-d spacetime metric. Observe that for some fixed $v$, $\{Z(v)\sin{u}\cos{w},Z(v)\sin{u}\sin{w},Z(v)\cos{u}\}$ forms a parametrisation of a 2-sphere with radius $Z(v)$. Ergo, the geometry of the 3-manifold of revolution constructed in this manner is that of having it being foliated by 2-spheres whose centres are on $\vec{\psi}$.

The Schwarzschild metric is a trivial application of this method by adding 2-spheres along a straight line (\cite{Vee2012,Vee2013}, section 3 of \cite{Vee2014}, or see section 2 below). Whilst Eq. (\ref{method}) gives a static or time-independent 3-manifold, it is not difficult to construct stationary or dynamical spacetimes. An explicit example of a dynamical spacetime can be found in Ref. \cite{Vee2013} where an inflating wormhole is constructed by letting $\vec{\psi}$ and $Z$ depend on time.

In this paper, we apply this method to perturbing the Schwarzschild metric in three separate ways:
\begin{enumerate}
\item
Non-rotating 2-spheres are added along a plane curve slightly deviated from the ``Schwarzschild line''. (In this paper, \emph{Schwarzschild line} refers to the straight line $\vec{\psi}$ where application of this method produces the usual Schwarzschild solution.)
\item
Non-rotating general topological 2-spheres are added along the Schwarzschild line.
\item
Slow-rotating 2-spheres are added along the Schwarzschild line.
\end{enumerate}
In other words, our goal here will be to obtain vacuum solutions (as opposed to allowing essentially uncontrolled freedom in the materials supporting traversable wormholes). We appeal to the use of perturbations around the Schwarzschild metric instead of solving the full vacuum field equations where for instance: 1) the plane curve is only slightly deviated from the Schwarzschild line; 2) the general topological 2-spheres are only slightly warped from those 2-spheres which produce the Schwarzschild metric; and 3) the 2-spheres are only slowly rotating; because the full non-perturbative equations are highly non-linear. More crucially, there is no guarantee for a general vacuum spacetime to admit foliations according to this prescription and thus there might not even exist such a vacuum solution.


Such 3-d spatial manifolds can obviously be foliated into 2-spheres (by construction). This provides a means of describing the (3+1)-d spacetime in a tractable manner in trying to solve the vacuum field equations, offering a useful geometrical structure and symmetry, in contrast to solving for a general manifold without much known additional structure. For example, it is easy to construct manifolds of revolution around a curve which are manifestly axially symmetric (explicitly independent of the $w$-coordinate in Eq. (\ref{method})). The notion of 3-d hypersurfaces assumed to be foliated by compact 2-surfaces in general has been employed elsewhere in general relativity apart from our consideration here to construct vacuum spacetimes (and originally for constructing curved traversable wormholes), though not based on the idea of generating manifolds of revolution around a curve. For instance, Frauendiener derived an integral formula on such hypersurfaces assumed to be foliated by compact 2-surfaces \cite{Fra97}. This has several nice applications, most notably in deriving the Bondi mass-loss formula for a general asymptotically flat spacetime, as well as the Penrose inequality for spherically symmetric spacetimes. In fact, a way to arrive at the Penrose inequality involves foliating an initial data hypersurface by 2-surfaces according to the inverse mean curvature flow \cite{Jang77,Ger73,fra01}. With this point of view, general relativity may be thought of as unique in the sense that one has this geometrical picture to regard the 3-space as being composed of individual layers of 2-surfaces and then try to solve the PDEs with such a priori structure in mind. A general set of PDEs does not possess such structures that could otherwise assist in trying to find solutions.

In general, this perturbative approach can be applied to any spacetime which is known to admit a decomposition into spheres being added along a curve. Introducing a curve slightly deviated from the actual one, adding slightly warped topological 2-spheres, or adding slow-rotating 2-spheres would allow one to study effects due to such departures. For the Schwarzschild metric, this is a way of breaking the spherical symmetry whilst still providing a means of keeping the analysis tractable.

This paper is organised according to the ordering of the three separate perturbations: In section 2, we discuss the first order vacuum solution due to a plane curve slightly deviated from the Schwarzschild line, and find that no higher order vacuum solution exists. We then discover and explain that this linearised solution is actually equivalent to a gauge transformation of the Schwarzschild metric via the Regge-Wheeler formalism \cite{ReggeWheeler,Sarbach}. The next section is devoted to the idea of adding general topological 2-spheres to the Schwarzschild line. Whilst we make no attempt to completely solve the most general resulting linearised vacuum field equations, we focus on the solutions for several special cases: in particular those where the 2-surfaces turn out to be non-compact (so they are not topological 2-spheres), giving rise to signature-changing spacetime metrics that contain closed timelike curves. Finally, we deal with the details of the slowly rotating 2-spheres in section 4, and also explore a non-perturbative generalisation to this method by adding rotating 2-ellipsoids. It turns out however, that adding rotating 2-ellipsoids does not lead to the full Kerr solution because the resulting form of the metric is only a subset of the most general ellipsoidal metric \cite{Kra,Ple}.

\section{Adding non-rotating 2-spheres to plane curve slightly deviated from the Schwarzschild line}

Consider the Schwarzschild exterior vacuum solution in Schwarzschild coordinates,
\begin{eqnarray}
ds^2=-\left(1-\frac{R_S}{r}\right)dt^2+\frac{1}{1-R_S/r}dr^2+r^2(d\theta^2+\sin^2{\theta}\ d\phi^2).
\end{eqnarray}
By the general method of constructing manifolds of revolution around a given curve, this form of the Schwarzschild metric is constructed by the radial function $Z(r)=r$ along the line $\vec{\psi}(r)=(0,0,0,f(r))$ in $\R^4$, where $f(r)=2\sqrt{R_S(r-R_S)}$. Note that since $\vec{\psi}$ is one of the axes of $\R^4$, the three orthonormal vectors $\hat{n}_1,\hat{n}_2,\hat{n}_3$ are just the three other standard Euclidean coordinate basis vectors for $\R^4$.

For a small deviation from the straight line $\vec{\psi}(r)=(0,0,0,f(r))$, let $\tilde{\psi}(r)=(0,0,\varepsilon g(r),f(r))$. The 3-manifold of revolution (embedded into $\R^4$) is constructed by adding the oscillatory terms $\{r\sin{\theta}\cos{\phi},r\sin{\theta}\sin{\phi},r\cos{\theta}\}$ respectively along three mutually orthonormal directions which are perpendicular to the tangent vector, i.e. $\hat{n}_1=(1,0,0,0)$, $\hat{n}_2=(0,1,0,0)$, $\hat{n}_3=(0,0,f'(r),-\varepsilon g'(r))/d$ where $d=\sqrt{f'(r)^2+\varepsilon^2g'(r)^2}\approx f'(r)$. The resulting 3-manifold of revolution is thus
\begin{eqnarray}
\vec{\sigma}(r,\theta,\phi)=(r\sin{\theta}\cos{\phi},r \sin{\theta}\sin{\phi},r\cos{\theta}+\varepsilon g(r),f(r)-\varepsilon g'(r)r\cos{\theta}/f'(r)).
\end{eqnarray}
It is known from the general method of constructing 3-manifolds of revolution around plane curves \cite{Vee2012,Vee2013} that $g_{rr}$ is a complicated term (even here for first order in $\varepsilon$.) If we consider instead the three orthonormal directions to be the $\hat{e}_1, \hat{e}_2, \hat{e}_3$ axes, perpendicular to the unperturbed Schwarzschild line $\vec{\psi}$ instead of $\tilde{\psi}$ \footnote{This similar to how the (2+1)-d and (3+1)-d helical wormholes were constructed by this freedom in choosing the orthonormal directions not to be perpendicular to the tangent vector to the given curve (there, the given curve is the helix, and a ``tube'' around the helix was constructed), giving rise to simpler equations \cite{Vee2012,Vee2013}.}, then the resulting 3-manifold of revolution is
\begin{eqnarray}\label{plane}
\vec{\sigma}(r,\theta,\phi)=\left(r\sin{\theta}\cos{\phi},r\sin{\theta}\sin{\phi},r\cos{\theta}+\varepsilon g(r),2\sqrt{R_S(r-R_S)}\right).
\end{eqnarray}
The partial derivatives are
\begin{eqnarray}
\vec{\sigma}_r&=&\left(\sin{\theta}\cos{\phi},\sin{\theta}\sin{\phi},\cos{\theta}+\varepsilon g'(r),\sqrt{\frac{R_S}{r-R_S}}\right)\\
\vec{\sigma}_\theta&=&\left(r\cos{\theta}\cos{\phi},r\cos{\theta}\sin{\phi},-r\sin{\theta},0\right)\\
\vec{\sigma}_\phi&=&\left(-r\sin{\theta}\sin{\phi},r\sin{\theta}\cos{\phi},0,0\right).
\end{eqnarray}
The spatial 3-metric (to first order in $\varepsilon$) using $g_{ij}=\vec{\sigma}_i\cdot\vec{\sigma}_j$ (see footnote [7]) is then
\begin{eqnarray}
ds^2=\left(\frac{1}{1-R_S/r}+2\varepsilon g'(r)\cos{\theta}\right)dr^2+r^2(d\theta^2+\sin^2{\theta}\ d\phi^2)-2\varepsilon rg'(r)\sin{\theta}\ drd\theta,
\end{eqnarray}
and the resulting (3+1)-d spacetime metric being
\begin{eqnarray}
ds^2&=&-\left(\left(1-\frac{R_S}{r}\right)+\varepsilon\chi(r,\theta)\right)dt^2+\left(\frac{1}{1-R_S/r}+2\varepsilon g'(r)\cos{\theta}\right)dr^2\nonumber\\&\ &+r^2(d\theta^2+\sin^2{\theta}\ d\phi^2)-2\varepsilon rg'(r)\sin{\theta}\ drd\theta,
\end{eqnarray}
since we assume the perturbed Schwarzschild spacetime to be static. The term $\chi(r,\theta)$ is the deviation from the Schwarzschild $g_{tt}$ corresponding to that of $g(r)$ on the Schwarzschild line. In a fully expanded solution, $g_{tt}$ can be written as $g_{tt}=-\left(\left(1-R_S/r\right)+\varepsilon\chi_{\varepsilon}+\varepsilon^2\chi_{\varepsilon^2}+\cdots\right)$.

We can proceed to calculate the inverse metric $g^{\mu\nu}$ from $g^{\mu\nu}g_{\nu\rho}=\delta^\mu_{\ \rho}$, the Christoffel symbols $\Gamma^{\mu}_{\alpha\beta}=\frac{1}{2}g^{\mu\nu}(g_{\alpha\nu,\beta}+g_{\beta\nu,\alpha}-g_{\alpha\beta,\nu})$, and finally the Ricci tensor $R_{\mu\nu}=\Gamma^{\lambda}_{\mu\nu,\lambda}-\Gamma^{\lambda}_{\mu\lambda,\nu}+\Gamma^{\lambda}_{\mu\nu}\Gamma^{\delta}_{\lambda\delta}-\Gamma^{\delta}_{\mu\lambda}\Gamma^{\lambda}_{\nu\delta}$. The non-zero components are:
\begin{eqnarray}
R_{tt}&=&\varepsilon\frac{R_S(r-R_S)\cos{\theta}}{2r^5}\left[(2r-R_S)g'-r(r-R_S)g''\right]\nonumber\\&\ &+\varepsilon\left[\frac{1}{2r^2}\chi_{\theta\theta}+\frac{1}{2r^2}\chi_\theta\cot{\theta}+\frac{(r-R_S)}{2r}\chi_{rr}+\frac{(4r-5R_S)}{4r^2}\chi_r+\frac{R_S^2}{4r^3(r-R_S)}\chi\right]\\
R_{rr}&=&\varepsilon\frac{R_S\cos{\theta}}{2r^3(r-R_S)}\left[(2r-3R_S)g'-3r(r-R_S)g''\right]\nonumber\\&\ &+\varepsilon\left[-\frac{r}{2(r-R_S)}\chi_{rr}+\frac{R_S}{4(r-R_S)^2}\chi_r-\frac{R_S(4r-3R_S)}{4r(r-R_S)^3}\chi\right]\\
R_{\theta\theta}&=&-\varepsilon\frac{R_S\cos{\theta}}{r^2}\left[R_S g'+r(r-R_S)g''\right]+\varepsilon\left[-\frac{r}{2(r-R_S)}\chi_{\theta\theta}-\frac{r}{2}\chi_r+\frac{R_S}{2(r-R_S)}\chi\right]\\
\frac{R_{\phi\phi}}{\sin^2{\theta}}&=&-\varepsilon\frac{R_S\cos{\theta}}{r^2}\left[R_S g'+r(r-R_S)g''\right]\nonumber\\&\ &+\varepsilon\left[-\frac{r}{2(r-R_S)}\chi_\theta\cot{\theta}-\frac{r}{2}\chi_r+\frac{R_S}{2(r-R_S)}\chi\right]\\
R_{r\theta}&=&R_{\theta r}=\varepsilon\frac{R_S\sin{\theta}}{2r^2}g'+\varepsilon\left[\frac{(2r-R_S)}{4(r-R_S)^2}\chi_\theta-\frac{r}{2(r-R_S)}\chi_{r\theta}\right].
\end{eqnarray}

Here is a useful relation by inspection:
\begin{eqnarray}
R_{\theta\theta}-\frac{R_{\phi\phi}}{\sin^2{\theta}}=-\varepsilon\frac{r}{2(r-R_S)}(\chi_{\theta\theta}-\chi_{\theta}\cot{\theta}).
\end{eqnarray}
For $R_{\mu\nu}=0$, this gives $\chi_{\theta\theta}=\chi_{\theta}\cot{\theta}$ which can be integrated to $\chi(r,\theta)=\chi_1(r)\cos{\theta}+\chi_2(r)$. Using $\chi_{\theta\theta}=\chi_{\theta}\cot{\theta}$, a second useful relation can be found:
\begin{eqnarray}
r(r-R_S)R_{rr}+\frac{r^3}{(r-R_S)}R_{tt}+2R_{\theta\theta}=\varepsilon\frac{2R_S\cos{\theta}}{r^2}[(r-2R_S)g'-2r(r-R_S)g''],
\end{eqnarray}
and $R_{\mu\nu}=0$ implies that $(r-2R_S)g'-2r(r-R_S)g''=0$, integrated to give $g(r)=A(r+2R_S)\sqrt{r-R_S}+B$. Again, using $\chi_{\theta\theta}=\chi_{\theta}\cot{\theta}$, there is a third useful relation:
\begin{eqnarray}
&\ &\frac{r(r-R_S)}{4}R_{rr}+\frac{r^3}{4(r-R_S)}R_{tt}-\frac{1}{2}R_{\theta\theta}-R_{r\theta}(r\cot{\theta})=\nonumber\\&\ &\varepsilon\left[\cot{\theta}\left(\frac{r^2}{2(r-R_S)}\chi_{r\theta}-\frac{R_S r}{4(r-R_S)^2}\chi_\theta\right)+\frac{r}{2}\chi_r-\frac{R_S}{2(r-R_S)}\chi\right],
\end{eqnarray}
where using $\chi(r,\theta)=\chi_1(r)\cos{\theta}+\chi_2(r)$ and $R_{\mu\nu}=0$ give
\begin{eqnarray}
\cos{\theta}\left(\frac{r^2}{2(r-R_S)}\chi_1'-\frac{R_S r}{4(r-R_S)^2}\chi_1\right)=\frac{r}{2}(\chi_1'\cos{\theta}+\chi_2')-\frac{R_S}{2(r-R_S)}(\chi_1\cos{\theta}+\chi_2).
\end{eqnarray}
Collecting terms without $\cos{\theta}$ leads to $\chi_2(r)=C(r-R_S)/r$, and collecting terms with $\cos{\theta}$ yields $\chi_1(r)=D\sqrt{r-R_S}/r$. It can be checked that with this $\chi$ and $g$, all components of the Ricci tensor $R_{\mu\nu}$ are identically zero if $D=AR_S$. Ergo,
\begin{eqnarray}
g(r)&=&A(r+2R_S)\sqrt{r-R_S}+B\\
\chi(r,\theta)&=&C\left(1-\frac{R_S}{r}\right)+\frac{AR_S}{r}\sqrt{r-R_S}\cos{\theta}
\end{eqnarray}
form the solution to the first order vacuum field equations for a slightly deviated plane curve from the Schwarzschild line. Thus, the sought after plane curve is
\begin{eqnarray}\label{planecurve}
\tilde{\psi}(r)=\left(0,0,\varepsilon A(r+2R_S)\sqrt{r-R_S},2\sqrt{R_S(r-R_S)}\right),
\end{eqnarray}
where the arbitrary constant $B$ is made zero by an appropriate choice for the origin of the coordinate system. The corresponding metric is
\begin{eqnarray}
ds^2&=&-\left((1+\varepsilon C)\left(1-\frac{R_S}{r}\right)+\varepsilon\frac{AR_S}{r}\sqrt{r-R_S}\cos{\theta}\right)dt^2\nonumber\\&\ &+\left(\frac{1}{1-R_S/r}+\frac{3\varepsilon Ar\cos{\theta}}{\sqrt{r-R_S}}\right)dr^2+r^2(d\theta^2+\sin^2{\theta}\ d\phi^2)-\frac{3\varepsilon Ar^2\sin{\theta}}{\sqrt{r-R_S}}\ drd\theta.
\end{eqnarray}

Next, we look into the second order vacuum field equations where the curve is $\tilde{\psi}(r)=(0,0,\varepsilon g(r)+\varepsilon^2 h(r),f(r))$, with $g$ being the solution that we just found, and $h$ being the second order perturbation. The details are worked out in the appendix A, showing that no second order solution exists. So, the above metric can be rewritten as
\begin{eqnarray}\label{solplane}
ds^2&=&-\left(1-\frac{R_S}{r}+\varepsilon\frac{R_S}{r}\sqrt{r-R_S}\cos{\theta}\right)dt^2+\left(\frac{1}{1-R_S/r}+\frac{3\varepsilon r\cos{\theta}}{\sqrt{r-R_S}}\right)dr^2\nonumber\\&\ &+r^2(d\theta^2+\sin^2{\theta}\ d\phi^2)-\frac{3\varepsilon r^2\sin{\theta}}{\sqrt{r-R_S}}\ drd\theta.
\end{eqnarray}
Note that we have gotten rid of $C$ by a rescaling of the time coordinate, following the absence of a higher order solution, and then absorbed $A$ into the perturbation strength, $\varepsilon$.

By adding 2-spheres in the application of this method of generating manifolds of revolution around a plane curve, we demand a rather high symmetry as compared to adding an arbitrary compact 2-surface. This is because only one degree of freedom defines a 2-sphere, viz. its radius. Whilst this high symmetry and lack of degree of freedom do lead to a unique linearised vacuum solution, it turns out to be overly restrictive against accommodating higher order Ricci flat spacetimes.

A possible resolution to this would be to drop the demand that the resulting spacetime be Ricci flat. This was in fact the original purpose of this method to represent curved traversable wormholes \cite{Vee2012,Vee2013}. It could also perhaps be adapted to construct interior solutions for non-spherically symmetric astrophysical objects. Our focus for this paper on the other hand, is the vacuum region outside a (slightly) non-spherically symmetric body which may be constructed using our method. Hence, an alternative approach in attempting to find higher order vacuum solutions (or eventually if possible, full vacuum solutions) would be to relax the use of the high symmetry of the sphere. A natural first step towards this generalisation would be to consider the ellipsoid, which has its eccentricity as an additional degree of freedom (which we shall be attempting in section 4 to see if this gives the full Kerr). The essential motivation for employing the sphere is its high symmetry leading to reasonably tractable equations in order to glean some useful insights, as we have presented. A lack of symmetry is accompanied by significantly more intricate technical details. Adding general topological 2-spheres will be the topic of the next section.

Meanwhile, let us try to better understand the metric given by Eq. (\ref{solplane}). Is this a new linearised vacuum solution? In the study of perturbations over a background metric only to first order, there exists the degree of freedom due to a gauge transformation. For the Schwarzschild metric in particular, the Regge-Wheeler formalism \cite{ReggeWheeler,Sarbach} originally developed to study stability questions may shed some light to possibly identify the nature of the spacetime described by Eq. (\ref{solplane}). By comparing the perturbation terms (i.e. those involving $\varepsilon$) to the general perturbation terms in Ref. \cite{ReggeWheeler} (which are their Eqs. (12) and (13)), it can be observed by inspection that our metric Eq. (\ref{solplane}) corresponds to the $l=1$ even-parity sector. This is known to be a gauge transformation of the Schwarzschild metric to first order \cite{Sarbach}. More specifically, the general gauge transformations are given in Section II B of Ref. \cite{Sarbach}. Using their notation, we have $H_{tt}=-\varepsilon R_S\sqrt{r-R_S}/r$, $H_{rr}=3\varepsilon r/\sqrt{r-R_S}$, $H_{tr}=0$, $Q_t=0$, $Q_r=3\varepsilon r^2/2\sqrt{r-R_S}$, $K=0$, with $G$ not present for $l=1$. Then, the following choice of $\xi_t=0$, $\xi_r=-\varepsilon r^2/\sqrt{r-R_S}$, $f=-\varepsilon\sqrt{r-R_S}$ would eliminate all the first order terms under the gauge transformation to leave the background Schwarzschild metric with zero perturbation \footnote{See acknowledgments.}.

\newpage

\section{Adding slightly warped topological 2-spheres to Schwarzschild line}

We would now add general topological 2-spheres along the Schwarzschild line, $\vec{\psi}(r)=(0,0,0,f(r))$, where $f(r)=2\sqrt{R_S(r-R_S)}$. This is done by adding the following oscillatory terms $\{(r+\varepsilon\rho_x(r,\theta,\phi))\sin{\theta}\cos{\phi},(r+\varepsilon\rho_y(r,\theta,\phi))\sin{\theta}\sin{\phi},(r+\varepsilon\rho_z(r,\theta,\phi))\cos{\theta}\}$ to the three coordinate basis vectors $\hat{e}_1, \hat{e}_2, \hat{e}_3$ to give the following 3-manifold of revolution:
\begin{eqnarray}\label{gentop}
\vec{\sigma}(r,\theta,\phi)&=&\left((r+\varepsilon\rho_x(r,\theta,\phi))\sin{\theta}\cos{\phi},(r+\varepsilon\rho_y(r,\theta,\phi))\sin{\theta}\sin{\phi},(r+\varepsilon\rho_z(r,\theta,\phi))\cos{\theta},\right.\nonumber\\&\ &\left.f(r)\right).
\end{eqnarray}
In order to ensure that these oscillatory terms represent compact topological 2-spheres, appropriate boundary conditions such as periodicity of $\phi$ over an interval of $2\pi$ and that $\rho_x,\rho_y,\rho_z$ do not become infinite should be imposed. We see below however, that there are certain classes of linearised vacuum solutions where the poles $\theta=0,\pi$ or the equator $\theta=\pi/2$ of these 2-surfaces would indeed blow up, and so we would not strictly rule out such solutions.

The 3-d spatial metric for Eq. (\ref{gentop}) can be calculated, and then appended to be a (3+1)-d spacetime metric with $g_{tt}=-(1-R_S/r+\varepsilon\Upsilon(r,\theta,\phi))$, $g_{ti}=0$. As was done in the previous section, the Ricci tensor is calculated to first order in $\varepsilon$ and would equal to zero for vacuum spacetimes. The most general case here however, is unwieldy and not very illuminating. Each non-zero component of $R_{\mu\nu}$ is of the order of a couple of pages long, after full simplification by an algebraic software like Mathematica. We therefore make no attempt here to completely find all solutions. Nevertheless, there are several classes of interesting vacuum solutions which we now present.

\newpage

\subsection{Adding displaced 2-spheres along the Schwarzschild line, which is equivalent to adding 2-spheres to plane curve}

If the ``warping'' of the 2-spheres are given by $\rho_x=\rho_y=0$, $\rho_z=g(r)\sec{\theta}$, where $g(r)=(r+2R_S)\sqrt{r-R_S}$, then the 3-manifold of revolution in Eq. (\ref{gentop}) becomes
\begin{eqnarray}
\vec{\sigma}(r,\theta,\phi)&=&\left(r\sin{\theta}\cos{\phi},r\sin{\theta}\sin{\phi},r\cos{\theta}+\varepsilon g(r),f(r)\right).
\end{eqnarray}
This leads to the resulting vacuum spacetime which is exactly Eq. (\ref{solplane}) found in the previous section. This association arises from $\rho_z=g(r)\sec{\theta}$ which becomes infinite at the equator $\theta=\pi/2$, though it precisely cancels out the $\cos{\theta}$ factor which multiplies it. Hence, what we see here is the relation between adding ``displaced'' 2-spheres along the Schwarzschild line (i.e. the 2-spheres $\{r\sin{\theta}\cos{\phi},r\sin{\theta}\sin{\phi},r\cos{\theta}\}$ get displaced by $\varepsilon g(r)\hat{e}_3$ when added to the Schwarzschild line), with adding 2-spheres directly to the plane curve $\vec{\psi}(r)=(0,0,\varepsilon g(r),f(r))$.

In fact more generally, take $\rho_x=\varepsilon_xg(r)\csc{\theta}\sec{\phi}$, $\rho_y=\varepsilon_yg(r)\csc{\theta}\csc{\phi}$, $\rho_z=\varepsilon_zg(r)\sec{\theta}$. Then, the 3-manifold of revolution is
\begin{eqnarray}
\vec{\sigma}(r,\theta,\phi)&=&(r\sin{\theta}\cos{\phi}+\varepsilon\varepsilon_xg(r),r\sin{\theta}\sin{\phi}+\varepsilon\varepsilon_yg(r),r\cos{\theta}+\varepsilon\varepsilon_zg(r),f(r)),
\end{eqnarray}
where here we have three independent perturbations of strengths $\varepsilon\varepsilon_x,\varepsilon\varepsilon_y,\varepsilon\varepsilon_z$ normalised by $\varepsilon_x^2+\varepsilon_y^2+\varepsilon_z^2=1$. This may alternatively be obtained by adding 2-spheres to the plane curve $\vec{\psi}(r)=(\varepsilon\varepsilon_xg(r),\varepsilon\varepsilon_yg(r),\varepsilon\varepsilon_zg(r),f(r))$. It can be shown that the resulting (3+1)-d linearised vacuum metric would be obtained by computing the above 3-d spatial metric and then appending the term $g_{tt}=-\left(1-R_S/r+\varepsilon R_S\sqrt{r-R_S}(\varepsilon_x\sin{\theta}\cos{\phi}+\varepsilon_y\sin{\theta}\sin{\phi}+\varepsilon_z\cos{\theta})/r\right)$. Of course, this is just a gauge transformation of the Schwarzschild metric, as explained towards the end of the last section. Nevertheless, it is appealing to relate such gauge transformations of the Schwarzschild metric with the geometrical viewpoint of adding 2-spheres to a plane curve or equivalently adding displaced 2-spheres to the Schwarzschild line.

\subsection{$\phi$-independence and $\rho_x=\rho_y$}

To glean possible linearised vacuum solutions, we focus here on the special cases where $\rho_x=\rho_y=\rho$, and all $\rho_x,\rho_y,\rho_z$, together with $\Upsilon$ are independent of $\phi$. The vacuum field equations are still rather long and regrettably not presented explicitly. It is perhaps much more instructive instead, for us to write down the solutions we found under the following situations \footnote{Even these solutions, Eqs. (\ref{genmetchan1}) and (\ref{genmetchan2}) are perhaps not the most general solutions since we assumed some way of splitting up terms into groups which go to zero independently, as opposed to requiring the entire thing go to zero. As an illustration, suppose $\alpha+\beta=0$. We assumed that $\alpha=\beta=0$, otherwise it would not have been obvious how the rest of the equations may be solved. Also, similar useful relations formed when solving for the plane curve case can also be formed here.} (1) $\rho_z=0$; (2) $\rho_x=\rho_y=\rho=0$. One can calculate the Ricci tensor for these solutions and find that they are indeed zero to first order in $\varepsilon$.

\subsubsection{$\rho_z=0$}

The general linearised vacuum solution for $\rho_z=0$ is:
\begin{eqnarray}\label{genmetchan1}
ds^2&=&-\left(1-\frac{R_S}{r}+\varepsilon\left(\frac{C_1R_S\sqrt{r-R_S}}{r}\cos{\theta}+\frac{C_3R_S\sqrt{r-R_S}}{r^2}+\left(\frac{C_4R_S}{r}+C_5\right)\left(1-\frac{R_S}{r}\right)\right)\right)dt^2\nonumber\\&\ &+\left(\frac{1}{1-R_S/r}+\varepsilon\left(\frac{3C_1r\cos{\theta}}{\sqrt{r-R_S}}+\frac{C_3}{\sqrt{r-R_S}}+C_4\right)\right)dr^2\nonumber\\&\ &+\left(r^2\sin^2{\theta}-2\varepsilon r\left((C_1g(r)+C_2)\cos{\theta}+\left(C_3\sqrt{r-R_S}+\frac{C_4}{2}(r-2R_S)\right)\cos^2{\theta}\right)\right)\csc^2{\theta}d\theta^2\nonumber\\&\ &+\left(r^2\sin^2{\theta}+2\varepsilon r\left((C_1g(r)+C_2)\cos{\theta}+\left(C_3\sqrt{r-R_S}+\frac{C_4}{2}(r-2R_S)\right)\right)\right)d\phi^2\nonumber\\&\ &+2\varepsilon\left(\frac{3C_1r^2\cos^2{\theta}}{2\sqrt{r-R_S}}-C_1g(r)-C_2-\frac{C_3(r-2R_S)\cos{\theta}}{2\sqrt{r-R_S}}+C_4R_S\cos{\theta}\right)\csc{\theta}drd\theta,
\end{eqnarray}
where $C_1,C_2,C_3,C_4,C_5$ are arbitrary constants.

This metric is the result of adding the 2-surfaces with oscillatory terms $\{(r+\varepsilon\rho(r,\theta))\sin{\theta}\cos{\phi},(r+\varepsilon\rho(r,\theta))\sin{\theta}\sin{\phi},r\cos{\theta}\}$ where $\rho=((C_1g(r)+C_2)\cos{\theta}+C_3\sqrt{r-R_S}+C_4(r-2R_S)/2))\csc^2{\theta}$ and $g(r)=(r+2R_S)\sqrt{r-R_S}$ is the same function $g(r)$ ($A=1$, $B=0$) appearing in the discussion of the plane curve deviation, to the Schwarzschild line and then having the $g_{tt}$ term appended to it. To facilitate our description of some remarkable features of this linearised vacuum solution, let us set the arbitrary constants $C_1=1$, $C_2=C_3=C_4=C_5=0$, to leave ourselves with:
\begin{eqnarray}
ds^2&=&-\left(1-\frac{R_S}{r}+\varepsilon\frac{R_S\sqrt{r-R_S}}{r}\cos{\theta}\right)dt^2+\left(\frac{1}{1-R_S/r}+\frac{3\varepsilon r\cos{\theta}}{\sqrt{r-R_S}}\right)dr^2\nonumber\\&\ &+\left(r^2\sin^2{\theta}-2\varepsilon rg(r)\cos{\theta}\right)\csc^2{\theta}d\theta^2+\left(r^2\sin^2{\theta}+2\varepsilon rg(r)\cos{\theta}\right)d\phi^2\nonumber\\&\ &+2\varepsilon\left(\frac{3r^2\cos^2{\theta}}{2\sqrt{r-R_S}}-g(r)\right)\csc{\theta}drd\theta,
\end{eqnarray}
with $\rho=g(r)\cos{\theta}\csc^2{\theta}$. We may further simplify this by the gauge transformation corresponding to subtracting off the perturbation terms in the $l=1$ even-parity sector, i.e. Eq. (\ref{solplane}) to arrive at:
\begin{eqnarray}
ds^2&=&-\left(1-\frac{R_S}{r}\right)dt^2+\frac{1}{1-R_S/r}dr^2+2\varepsilon\left(rg'(r)-g(r)\right)\csc{\theta}drd\theta\nonumber\\&\ &+\left(r^2\sin^2{\theta}-2\varepsilon rg(r)\cos{\theta}\right)\csc^2{\theta}d\theta^2+\left(r^2\sin^2{\theta}+2\varepsilon rg(r)\cos{\theta}\right)d\phi^2.
\end{eqnarray}
One can of course directly calculate the Ricci tensor for this metric to find that it is indeed zero to first order in $\varepsilon$.

\begin{figure}
\centering
\includegraphics[width=12cm]{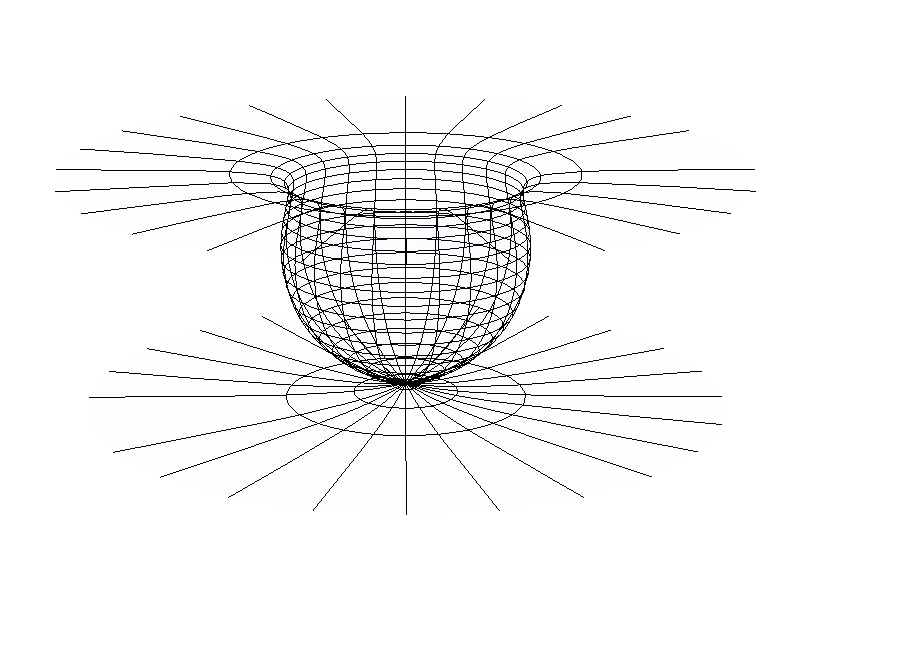}
\caption{The 2-surface which blows up at the poles.}
\end{figure}

This metric has unusual properties, which lies beyond the Regge-Wheeler formalism \cite{ReggeWheeler}. Firstly, it blows up at the poles $\theta=0,\pi$ due to the $\csc{\theta}$ terms. This is the result of adding the non-compact 2-surfaces where $\rho=g(r)\cos{\theta}\csc^2{\theta}$ becomes infinite there, see Fig. 1. Nevertheless, this is a coordinate singularity which can be eliminated by the coordinate transformation $e^q=\tan{(\theta/2)}$ where $q\in\R$ to give:
\begin{eqnarray}\label{sigchange}
ds^2&=&-\left(1-\frac{R_S}{r}\right)dt^2+\frac{1}{1-R_S/r}dr^2+2\varepsilon\left(rg'(r)-g(r)\right)drdq\nonumber\\&\ &+\left(r^2\sech^2{q}+2\varepsilon rg(r)\tanh{q}\right)dq^2+\left(r^2\sech^2{q}-2\varepsilon rg(r)\tanh{q}\right)d\phi^2.
\end{eqnarray}
In this form, this metric does not contain any term which go to infinity (for $r>R_S$). It does however, become degenerate and even changes its signature. A way to see this quickly is to recall that when $q\rightarrow\pm\infty$ (i.e. heading towards the poles), then $\sech{q}\rightarrow0$ and $\tanh{q}\rightarrow\pm1$. Consequently, either one of $g_{qq}$ or $g_{\phi\phi}$ would change its sign from $+$ to $-$, and the signature of the spacetime goes from $(1,3)$ to $(2,2)$. The region near the equator $q=0$ is where its signature remains as $(1,3)$.

In the reduction to $R_S=0$, the metric Eq. (\ref{sigchange}) becomes
\begin{eqnarray}\label{Min1}
ds^2&=&-dt^2+dr^2+\left(r^2\sech^2{q}+2\varepsilon r^{5/2}\tanh{q}\right)dq^2+\left(r^2\sech^2{q}-2\varepsilon r^{5/2}\tanh{q}\right)d\phi^2\nonumber\\&\ &+\varepsilon r^{3/2}drdq.
\end{eqnarray}
The signature-changing feature still persists as in the Schwarzschild case. It can be checked that this is Riemann flat to first order and is thus Minkowski spacetime by adding those 2-surfaces to the point $\vec{\psi}(r)=(0,0,0,0)$ (since $f(r)=2\sqrt{R_S(r-R_S)}=0$). (Note that the perturbation terms can be eliminated by a gauge transformation.)

Signature-changing metrics have been a lively subject in the literature \cite{metchange1,metchange2,metchange3}, with a recent work by Ref. \cite{metchange3} discussing how this may be a challenge for quantum gravity. Whilst Ref. \cite{metchange1} provided examples of continuous and discontinuous signature changes from $(1,3)$ to $(0,4)$, what we have here on the other hand is a way of constructing signature change from $(1,3)$ to $(2,2)$ (or it may also be adapted to go from $(0,4)$ to $(1,3)$ if $g_{tt}=1-R_S/r$ in Eq. (\ref{sigchange})) by adding non-compact 2-surfaces which blow up at the poles, and requiring that it is Ricci flat to first order. Notice that $\varepsilon$ can be as small as desired so that higher order terms can essentially be ignored, but this feature always remains as long as $\varepsilon\neq0$ because the polar angular coordinate is ``non-compact''. Hence, it is not necessary to look for higher order solutions.

\subsubsection{$\rho_x=\rho_y=\rho=0$}

\begin{figure}
\centering
\includegraphics[width=8cm]{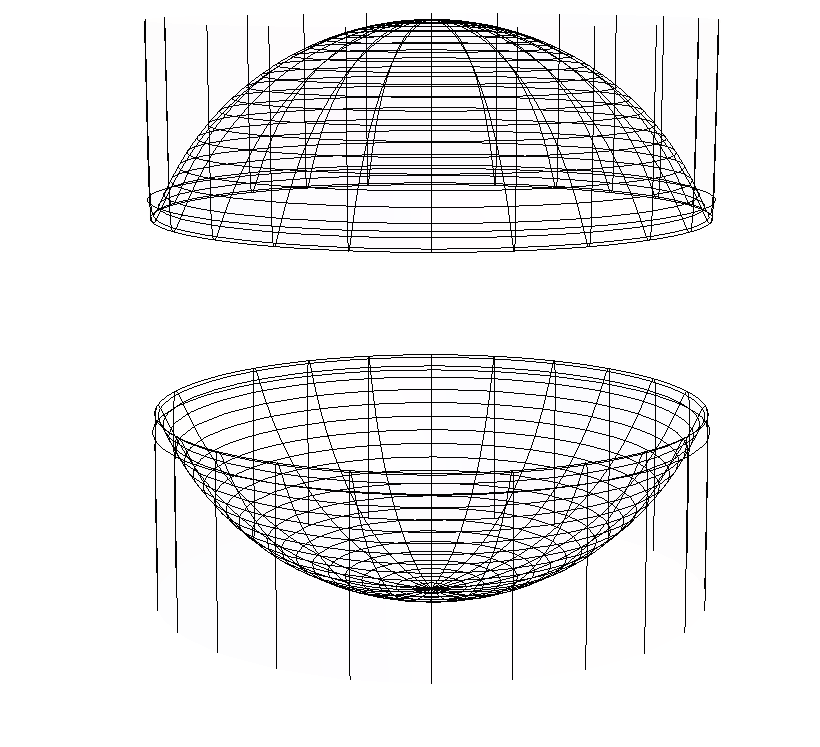}
\caption{The 2-surface which blows up at the equator.}
\end{figure}

The general linearised vacuum solution for {$\rho_x=\rho_y=\rho=0$ is:
\begin{eqnarray}\label{genmetchan2}
ds^2&=&-\left(1-\frac{R_S}{r}+\varepsilon\left(\frac{D_3R_S\sqrt{r-R_S}}{r^2}+\left(\frac{D_4R_S}{r}+D_5\right)\left(1-\frac{R_S}{r}\right)\right)\right)dt^2\nonumber\\&\ &+\left(\frac{1}{1-R_S/r}+\varepsilon\left(\frac{D_3}{\sqrt{r-R_S}}+D_4\right)\right)dr^2\nonumber\\&\ &+\left(r^2\cos^2{\theta}-2\varepsilon r\left(\left(D_3\sqrt{r-R_S}+\frac{D_4}{2}(r-2R_S)\right)\sin^2{\theta}\right)\right)\sec^2{\theta}d\theta^2\nonumber\\&\ &+r^2\sin^2{\theta}d\phi^2+2\varepsilon\left(\frac{D_3(r-2R_S)}{2\sqrt{r-R_S}}-D_4R_S\right)\sin{\theta}\sec{\theta}drd\theta,
\end{eqnarray}
where $D_3,D_4,D_5$ are arbitrary constants. (There are terms involving $D_1$ and $D_2$ analogous to $C_1$ and $C_2$ for the $\rho_z=0$ case. These however, correspond to the plane curve solution Eq. (\ref{solplane}) which is just a gauge transformation.) This metric is the result of adding the 2-surfaces with oscillatory terms $\{r\sin{\theta}\cos{\phi},r\sin{\theta}\sin{\phi},(r+\varepsilon\rho_z(r,\theta))\cos{\theta}\}$ where $\rho_z=(D_3\sqrt{r-R_S}+D_4(r-2R_S)/2))\sec^2{\theta}$, to the Schwarzschild line and then having the $g_{tt}$ term appended to it. Let us now focus on $D_3=D_5=0$, $D_4=1$ for the sake of discussion:
\begin{eqnarray}
ds^2&=&-\left(1+\varepsilon\frac{R_S}{r}\right)\left(1-\frac{R_S}{r}\right)dt^2+\left(\frac{1}{1-R_S/r}+\varepsilon\right)dr^2\nonumber\\&\ &+\left(r^2\cos^2{\theta}-\varepsilon r(r-2R_S)\sin^2{\theta}\right)\sec^2{\theta}d\theta^2+r^2\sin^2{\theta}d\phi^2-2\varepsilon R_S\sin{\theta}\sec{\theta}drd\theta,\ \ \ 
\end{eqnarray}
with $\rho_z(r,\theta)=0.5(r-2R_S)\sec^2{\theta}$ (see Fig. 2). This metric also lies beyond the Regge-Wheeler formalism \cite{ReggeWheeler} due to the blowing up at the equator $\theta=\pi/2$ of $\sec{\theta}$. Under the coordinate transformation $e^q=\tan{\theta}+\sec{\theta}$ where $q\in[0,\infty)$,
\begin{eqnarray}\label{sigchange2}
ds^2&=&-\left(1+\varepsilon\frac{R_S}{r}\right)\left(1-\frac{R_S}{r}\right)dt^2+\left(\frac{1}{1-R_S/r}+\varepsilon\right)dr^2\nonumber\\&\ &+\left(r^2\sech^2{q}-\varepsilon r(r-2R_S)\tanh^2{q}\right)dq^2+r^2\tanh^2{q}d\phi^2-2\varepsilon R_S\tanh{q}drdq.
\end{eqnarray}
Once again, we see that for $\varepsilon>0$ and as $q\rightarrow\infty$ (towards the equator), then $g_{qq}$ changes sign so that the metric changes signature from $(1,3)$ to $(2,2)$. On the other hand, if $\varepsilon<0$ and is reasonably close to zero (so that $g_{tt}$ and $g_{rr}$ do not change signs), then its signature remains as $(1,3)$. This coordinate system however, only covers the upper half of the equator from $\theta=0$ to $\pi/2$. The other half below the equator can be covered by the coordinate transformation $e^{-q}=-\tan{\theta}-\sec{\theta}$ where $q\in(-\infty,0]$, giving the same form of the metric as Eq. (\ref{sigchange2}). In other words, $\theta\in[0,\pi/2)$ corresponds to $q\in[0,\infty)$ by $e^q=\tan{\theta}+\sec{\theta}$, and $\theta\in(\pi/2,\pi]$ corresponds to $q\in(-\infty,0]$ by $e^{-q}=-\tan{\theta}-\sec{\theta}$.

In the limit where $R_S\rightarrow0$, we have
\begin{eqnarray}\label{Min2}
ds^2&=&-dt^2+\left(1+\varepsilon\right)dr^2+\left(r^2\sech^2{q}-\varepsilon r^2\tanh^2{q}\right)dq^2+r^2\tanh^2{q}d\phi^2,
\end{eqnarray}
which still has the signature-changing feature. This is Riemann flat to first order in $\varepsilon$, implying that it is Minkowski spacetime.

\section{Slow-rotating linearised vacuum solution}

Consider the following time-dependent 3-manifold of revolution around a straight line embedded into a 4-d Euclidean space:
\begin{eqnarray}\label{rotmanifold}
\vec{\sigma}(t,r,\theta,\phi)=(r\sin{\theta}\cos{(\phi+\omega\Omega(r,\theta)t)},r\sin{\theta}\sin{(\phi+\omega\Omega(r,\theta)t)},r\cos{\theta},f(r)),
\end{eqnarray}
where $t$ is the time coordinate as measured by a faraway observer, $r$, $\theta$ and $\phi$ being the usual spherical coordinates for 3-d Euclidean space, $f(r)=2\sqrt{R_S(r-R_S)}$, and $\omega$ is a constant.

The time-dependence built in here by replacing $\phi\rightarrow \phi+\omega\Omega(r,\theta)t$ from the static version in Eq. (\ref{plane}) with $\varepsilon=0$ (i.e. the 3-manifold of revolution for the Schwarzschild solution) to yield Eq. (\ref{rotmanifold}) represents the fact that the added spherical shells are rotating about the fourth coordinate axis (or the line $\vec{\psi}(r)=(0,0,0,f(r))$) with angular velocity $\omega\Omega(r,\theta)$, Fig. 3. This can be seen by choosing any particular point on the 3-manifold i.e. fixing some values of $r,\theta,\phi$, and noting that as time $t$ evolves this point would be rotated by an angle of $\omega\Omega(r,\theta)t$ about $\vec{\psi}$. The $r$-dependence on $\Omega$ implies that spheres of different radii $r$ may in general be rotating with different angular velocities. Its $\theta$-dependence allows for the possibility that a sphere of radius $r$ may be composed of rings of various latitude angles $\theta$ rotating with different angular velocities. This is a construction of spacetime itself that is rotating around $\vec{\psi}$ (as opposed to what one may usually consider instead, namely a rotating source of mass-energy). In other words, we are geometrically constructing the frame-dragging effect outside a rotating body.

\begin{figure}
\centering
\includegraphics[width=10cm]{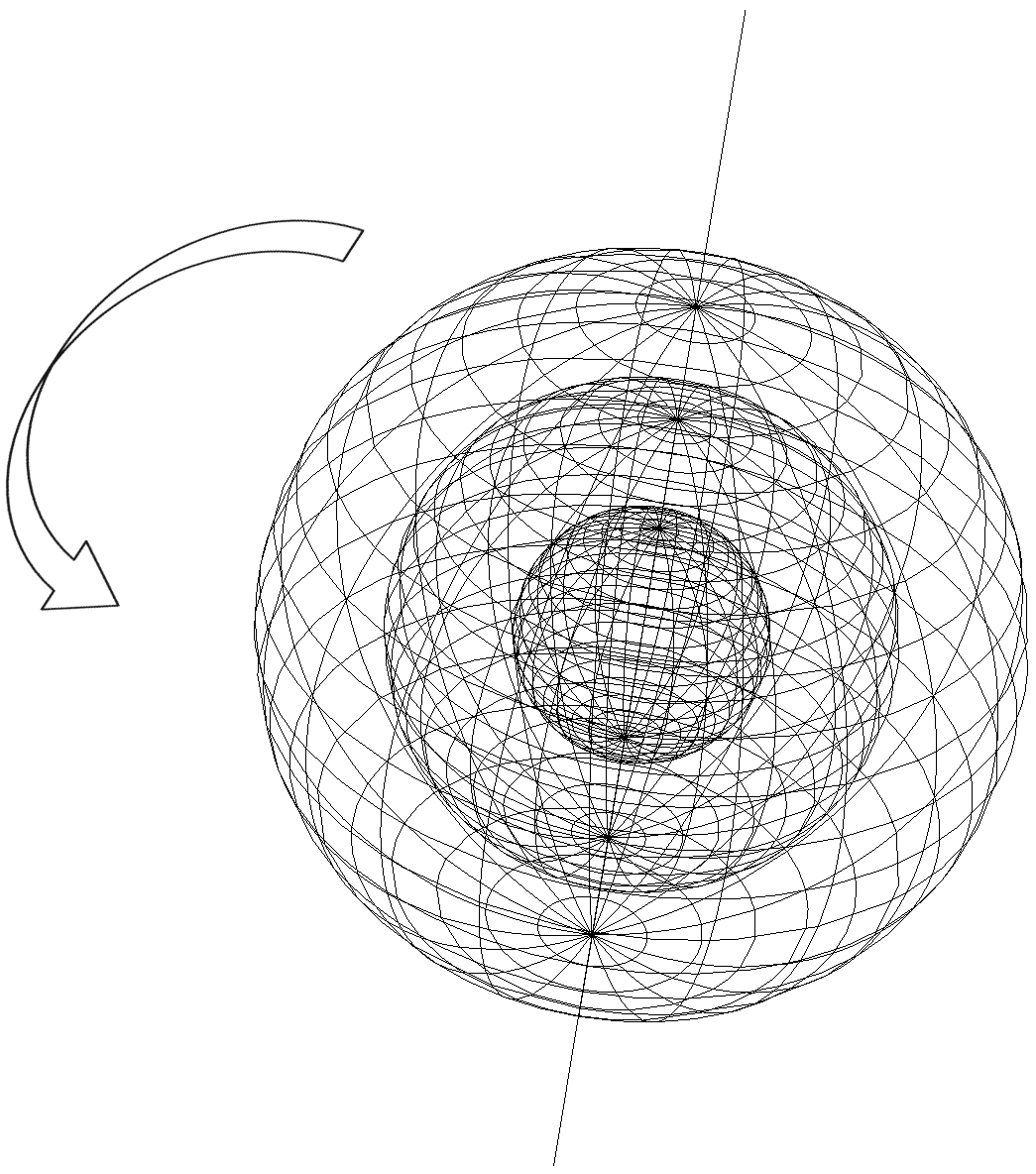}
\caption{Each shell is rotating with angular velocity $\omega\Omega(r,\theta)$, so in general each shell of radius $r$ is composed of rings of latitude $\theta$ that are rotating at different angular velocities.}
\end{figure}

The spatial metric for the 3-manifold given by Eq. (\ref{rotmanifold}) can be computed to give:
\begin{eqnarray}
ds^2&=&\left(\frac{1}{1-R_S/r}+\omega^2\Omega_r(r,\theta)^2t^2r^2\sin^2{\theta}\right)dr^2+\left(r^2+\omega^2\Omega_\theta(r,\theta)^2t^2r^2\sin^2{\theta}\right)d\theta^2\nonumber\\
&\ &+r^2\sin^2{\theta}\ d\phi^2+2\omega^2\Omega_r(r,\theta)\Omega_\theta(r,\theta)t^2r^2\sin^2{\theta}\ drd\theta+2\omega\Omega_r(r,\theta)tr^2\sin^2{\theta}\ drd\phi\nonumber\\
&\ &+2\omega\Omega_\theta(r,\theta)tr^2\sin^2{\theta}\ d\theta d\phi\\
&=&\frac{1}{1-R_S/r}dr^2+r^2d\theta^2+r^2\sin^2{\theta}(\omega\Omega_r(r,\theta)tdr+\omega\Omega_\theta(r,\theta)td\theta+d\phi)^2.
\end{eqnarray}
We extend this to the (3+1)-d spacetime metric:
\begin{eqnarray}\label{rotfull}
ds^2&=&-\left(1-\frac{R_S}{r}+\omega\Lambda\right)dt^2+\frac{1}{1-R_S/r}dr^2+r^2d\theta^2\nonumber\\
&\ &+r^2\sin^2{\theta}(\omega\Omega_r(r,\theta)tdr+\omega\Omega_\theta(r,\theta)td\theta+d\phi)^2,
\end{eqnarray}
so that $\omega=0$ corresponds to the Schwarzschild solution, and $\Lambda$ here represents the corresponding deviation from the Schwarzschild $g_{tt}$. As it is, it appears that this spacetime may be dynamical due to the explicit time dependence in the metric components. Upon the coordinate transformation $\phi'=\phi+\omega\Omega(r,\theta)t$ however \footnote{In the construction of this rotating manifold of revolution, we effectively replaced $\phi$ by $\phi+\omega\Omega(r,\theta)t$. This is not a coordinate transformation, since the resulting manifold which is composed of rotating shells is not the same as the original Schwarzschild one which is static. Upon extending to the (3+1)-d spacetime in Eq. (\ref{rotfull}) on the other hand, $\phi'=\phi+\omega\Omega(r,\theta)t$ is indeed a coordinate transformation.}, we see that $d\phi'=d\phi+\omega\Omega_r(r,\theta)tdr+\omega\Omega_\theta(r,\theta)td\theta+\omega\Omega(r,\theta)dt$ so that the spacetime metric (dropping the prime on $\phi'$) becomes
\begin{eqnarray}
ds^2&=&-\left(1-\frac{R_S}{r}+\omega\Lambda-\omega^2\Omega(r,\theta)^2r^2\sin^2{\theta}\right)dt^2+\frac{1}{1-R_S/r}dr^2+r^2d\theta^2+r^2\sin^2{\theta}d\phi^2\nonumber\\&\ &-2\omega\Omega(r,\theta)r^2\sin^2{\theta}dtd\phi,
\end{eqnarray}
or to first order in $\omega$,
\begin{eqnarray}\label{solrot}
ds^2&=&-\left(1-\frac{R_S}{r}+\omega\Lambda\right)dt^2+\frac{1}{1-R_S/r}dr^2+r^2d\theta^2+r^2\sin^2{\theta}d\phi^	2\nonumber\\&\ &-2\omega\Omega(r,\theta)r^2\sin^2{\theta}dtd\phi.
\end{eqnarray}
We can then proceed to solve the linearised vacuum field equations, as was done previously. Instead, we note that this is precisely the slow-rotating approximation to the Kerr solution \cite{Hobson} if $\Lambda=0$, $\Omega=2/r^3$ and $\omega$ is interpreted as the angular momentum (in appropriate units where $G=c=1$). Each shell of radius $r$ therefore rotates rigidly.

The frame-dragging effect is a characteristic of spacetimes that have a non-zero $g_{t\phi}$ metric component \cite{Hobson}. The construction we present above begins by generating a spacelike rotating 3-manifold of revolution around the line $\vec{\psi}$, which certainly gives a 3-d spatial metric that (by definition) involves no $g_{t\mu}$ components. We nevertheless obtain spatial cross-terms that depend explicitly on time $t$ (well, every term of the metric depends on $t$ except for $g_{\phi\phi}$), which is the geometrical manifestation of the rotating shells of the 3-manifold of revolution. The extension to the (3+1)-d spacetime metric by appending the term $g_{tt}=-(1-R_S/r+\omega\Lambda)$, $g_{ti}=0$ is justified by the fact that this is a slow-rotating perturbation of the Schwarzschild metric. Finally, it is the coordinate transformation $\phi'=\phi+\omega\Omega(r,\theta)t$ that fittingly eliminates all spatial cross-terms, leaving $g_{t\phi}$ as the only cross-term (and simultaneously removes all explicit $t$-dependence of the metric so that it becomes a stationary spacetime - the slow-rotating Kerr, in fact). We have thus shown that our rotating-shells construction is indeed equivalent to the usual notion of frame-dragging defined by the non-zero $g_{t\phi}$.


\subsection{Adding rotating ellipsoids}

In section 2, we discussed that the high symmetry of the sphere corresponds to a lack of degree of freedom. Consequently, it uniquely determines a linearised vacuum solution but admits no higher order vacuum solution. For the case of rotating shells where the first order solution is the slow-rotating Kerr metric, it is intriguing to explore if it is possible to actually arrive at the full Kerr solution. The Kerr is certainly not spherically symmetric, but is in fact an ellipsoidal spacetime \cite{Kra,Ple,Nar}. According to Ref. {\cite{Kra}}, an ellipsoidal vacuum spacetime has a congruence of local rest spaces with metric of the form:
\begin{eqnarray}\label{ellip}
ds^2=A(r,\theta)dr^2+(r^2+a^2\cos^2{\theta})d\theta^2+(r^2+a^2)\sin^2{\theta}d\phi^2,
\end{eqnarray}
where $A(r,\theta)>0$. Ref. \cite{Kra} then managed to obtain the Kerr solution in Boyer-Lindquist coordinates.

Now if we add the rotating ellipsoids $\{k(r)\sin{\theta}\cos{(\phi+\omega\Omega(r,\theta)t)},k(r)\sin{\theta}\sin{(\phi+\omega\Omega(r,\theta)t)},r\cos{\theta}\}$, where $k(r)^2=r^2+h^2$ ($h$ is a constant) to the line $\vec{\psi}(r)=(0,0,0,z(r))$, the resulting 3-manifold of revolution would have the following metric:
\begin{eqnarray}
ds^2&=&\left(z'(r)^2+\frac{r^2+h^2\cos^2{\theta}}{r^2+h^2}\right)dr^2+(r^2+h^2\cos^2{\theta})d\theta^2\nonumber\\&\ &+(r^2+h^2)\sin^2{\theta}(\omega\Omega_r(r,\theta)tdr+\omega\Omega_\theta(r,\theta)td\theta+d\phi)^2,
\end{eqnarray}
so the spacetime metric would be
\begin{eqnarray}
ds^2&=&g_{tt}dt^2+\left(z'(r)^2+\frac{r^2+h^2\cos^2{\theta}}{r^2+h^2}\right)dr^2+(r^2+h^2\cos^2{\theta})d\theta^2\nonumber\\&\ &+(r^2+h^2)\sin^2{\theta}(\omega\Omega_r(r,\theta)tdr+\omega\Omega_\theta(r,\theta)td\theta+d\phi)^2.
\end{eqnarray}
With the change of coordinates $\phi'=\phi+\omega\Omega(r,\theta)t$, so $d\phi'=d\phi+\omega\Omega_r(r,\theta)tdr+\omega\Omega_\theta(r,\theta)td\theta+\omega\Omega(r,\theta)dt$, the spacetime metric (dropping the prime on $\phi'$) becomes
\begin{eqnarray}
ds^2&=&\left(g_{tt}+\omega^2\Omega(r,\theta)^2(r^2+h^2)\sin^2{\theta}\right)dt^2+\left(z'(r)^2+\frac{r^2+h^2\cos^2{\theta}}{r^2+h^2}\right)dr^2\nonumber\\&\ &+(r^2+h^2\cos^2{\theta})d\theta^2+(r^2+h^2)\sin^2{\theta}d\phi^2-2\omega\Omega(r,\theta)(r^2+h^2)\sin^2{\theta}dtd\phi.
\end{eqnarray}
We see that our metric is of the ellipsoidal form in Eq. (\ref{ellip}) for $t=$ constant, where $A(r,\theta)=z'(r)^2+(r^2+h^2\cos^2{\theta})/(r^2+h^2)$ and $a=h$. To get the Kerr metric, one may try to match this with its Boyer-Lindquist form (Eq. (1.57) in Ref. \cite{VisserKerr}), where the $g_{\theta\theta}$ terms are identical. However, our metric is not general enough to give rise to a Ricci flat spacetime. In particular, the $g_{rr}$ term for the Kerr in Boyer-Lindquist coordinates is $g_{rr, BL}=(r^2+h^2\cos^2{\theta})/(r^2-2mr+h^2)$. No $z(r)$ can give this since it cannot be a function of $\theta$. (Recall that $z(r)$ is a parametrisation of the straight line $\vec{\psi}(r)=(0,0,0,z(r))$ and must only depend on one parameter, since the line is one-dimensional.)

Our construction only gives a subset of the most general form of the ellipsoidal metric, Eq. (\ref{ellip}). With this being unable to produce a Ricci flat rotating spacetime, it can nevertheless be adapted to build a rotating spacetime with matter-energy fields present. Can this be used to derive an interior solution to the Kerr spacetime? Or perhaps it may be an alternative way to construct rotating traversable wormholes \cite{Teo}.

\section{Discussion}

In this paper, we have formulated a general approach for a perturbative analysis on spacetimes that can be constructed by generating manifolds of revolution around a curve, and applied it to the Schwarzschild metric to explicitly demonstrate its utility. Whilst we developed this method via perturbations on the Schwarzschild metric, the bigger goal is of course to obtain a fully non-perturbative construction by generating manifolds of revolution along an arbitrary curve (under suitable differentiability conditions) where the curve and resulting manifold of revolution may be embedded into $\R^n$ with $n\geq4$. Although we have shown here that the first order vacuum solution by adding 2-spheres along a plane curve deviated from the Schwarzschild line is just a gauge transformation and no higher order solution exists, it must be emphasised that we chose the three orthonormal vectors to be perpendicular to the Schwarzschild line instead of the plane curve itself. This has certainly brought about valuable understanding towards this perturbative method, where the equations were reasonably tractable (so we could present the entire derivation explicitly for this, but not for the solutions in section 3 which are significantly more arduous) \footnote{We remark that the linearised vacuum equations for the plane curve were originally calculated and solved entirely by hand, before later on verified using Mathematica.}, and it can also be directly compared to adding ``displaced 2-spheres'' which we discussed in section 3. The knowledge of that plane curve deviation being just a gauge transformation also became salutary as it prompted an obvious simplification of the solutions in section 3 by subtracting off those perturbations. (See also the last sentence in footnote [19].) The question of whether there may exist vacuum solutions due to generating manifolds of revolution around a plane curve with the three orthonormal vectors being perpendicular to the plane curve itself or even with some general orientation is therefore still open.

One may recall that the analytic extension of the Kerr metric into $r<0$ (in Boyer-Lindquist coordinates) contains closed timelike curves \cite{Hobson}. This happens because the $g_{\phi\phi}$ component of the Kerr metric changes sign for small negative values of $r$, and hence the corresponding closed curve given by $\phi\in(-\pi,\pi], \theta=\pi/2$, with constant $t$ and $r$, becomes timelike. We see that something similar occurs also to the signature-changing metric of Eq. (\ref{sigchange}), though not for that of Eq. (\ref{sigchange2}). For the former, set $\varepsilon>0$ and consider any constant values for $t, r, q_\infty$, where $r>R_S$, and $q_\infty<\infty$ being sufficiently large so that $r^2\sech^2{q_\infty}-2\varepsilon rg(r)\tanh{q_\infty}<0$. Such a $q_\infty$ must exist since given any fixed $r$, we have that $r^2\sech^2{q}\rightarrow0$ and $2\varepsilon rg(r)\tanh{q}\rightarrow2\varepsilon r g(r)>0$, as $q\rightarrow\infty$. Consequently, the closed curve given by $\phi\in(-\pi,\pi]$ with constant $t, r, q_\infty$ is timelike. The latter on the other hand only has a signature change due to the metric component $g_{qq}$ changing sign. The curve given by $q\in[0,\infty)$ with constant $t, r, \phi$ is not closed, however. To obtain a closed curve, one may try to attribute it to a compact coordinate (in this case $\phi$), when the other coordinates are fixed. But since the $g_{\phi\phi}$ metric component here does not become negative, such closed curves are not timelike.

Unlike the Kerr metric where these closed timelike curves occur in the analytic extension for $r<0$ and can be hidden behind an event horizon, those of Eq. (\ref{sigchange}) occur at the ``poles'' corresponding to $\theta\approx0$ for any $r>R_S$, i.e. they exist essentially throughout the entire spacetime near $\theta=0$. This raises a highly intriguing question of how it may be physically possible to have such a family of closed timelike curves over an unbounded region of spacetime, especially when this is vacuum and not within an event horizon. Is this perturbation induced by some kind of exotic matter fields on the Schwarzschild mass of the star/singularity at $r=0$, or is this yet another gauge transformation of the Schwarzschild metric itself?

This time the nature of the metrics given by Eqs. (\ref{sigchange}) and (\ref{sigchange2}) are not so clear, since the Regge-Wheeler formalism does not apply, and there does not seem to be an obvious way to write down such gauge transformations. Whilst the Riemann tensor for these metrics can be calculated to show that they do contain some terms first order to $\varepsilon$, this does not necessarily imply that it is not a gauge transformation because the Riemann tensor for Eq. (\ref{solplane}), which is equivalent to the Schwarzschild metric (shown via the Regge-Wheeler formalism), also contains terms first order to $\varepsilon$. This therefore leaves an open question on how to extend the Regge-Wheeler formalism to deal with more general perturbations of the Schwarzschild metric to: 1) include our construction method developed here; or otherwise 2) concoct an entirely new formalism. Note that we have the more general first order vacuum solutions given by Eqs. (\ref{genmetchan1}) and (\ref{genmetchan2}) which may possess even more unusual properties; or a considerable future goal would be to show that they turn out as being diffeomorphic to the Schwarzschild metric.

Even if this is ultimately demonstrated to be just a gauge transformation, this method may nevertheless be alternatively viewed as a way to construct signature-changing metrics, achieved by blowing up the originally compact topological 2-spheres. In particular, the reduction to Minkowski spacetime, Eq. (\ref{Min1}), is in coordinates where the signature of flat spacetime (to first order) changes, giving rise to such a phenomenon of closed timelike curves (since $g_{\phi\phi}$ for Eq. (\ref{Min1}) changes sign). The explanation is of course that the spacetime is not absolutely flat, but does actually contain higher order mass-energy fields, which would perhaps violate the standard energy conditions in order to support the closed timelike curves \cite{CTC}. It is certainly worth keeping in mind that such required exotic matter can be infinitesimal, since $\varepsilon\neq0$ may be made arbitrarily small with the spacetime always satisfying the linearised vacuum field equations whilst preserving the signature-changing property and the existence of the closed timelike curves.

\section{Concluding remarks}

To summarise, we carried out three different types of perturbations: by introducing a slight deviation to the curve, adding warped topological 2-spheres, and adding rotating 2-spheres instead of non-rotating ones. Applying these to the Schwarzschild metric, the first turns out not to be a new solution, but just a gauge transformation. Adding general topological 2-spheres to the Schwarzschild line leads to a raft of possible solutions, where we only studied several special cases here. Perhaps the most intriguing of such vacuum solutions are those whose signatures change via the ``blowing up'' of the originally compact topological 2-spheres, representing a new class of vacuum solutions containing closed timelike curves. Finally, the third is a geometrical construction of the frame-dragging phenomenon which produces the slow-rotating approximation of the Kerr spacetime. This is arguably a more elegant geometrical derivation compared to the much longer gravitomagnetic one. It also does not require one to already know the full Kerr solution (to carry out linearisation) whose derivation can become quite technically involved \footnote{Although the full Kerr solution is already known \cite{Kerr,VisserKerr} and then linearisation to slow-rotation can easily be obtained from it, its original derivation came about via advanced techniques involving the notions of Petrov classification, algebraically special spacetimes, and the Goldberg-Sachs theorem - not something that are typically included within a basic general relativity course. Brute force derivation of the Kerr metric is by no means a simple task either \cite{VisserKerr}. The approach presented in this paper based on a geometrical construction is perhaps of strong appeal, since a person new to general relativity would be able to arrive at the slow-rotating metric with minimal demand on specialised tools.}.

Overall, this method can be used to study such perturbation effects on other known exact solutions of the Einstein field equations, if those spacetimes can be decomposed in such a manner (for example, the Schwarzschild-de Sitter spacetime). This includes both the exterior vacuum region $R_{\mu\nu}=0$ as well as interior solutions where $G_{\mu\nu}=R_{\mu\nu}-g_{\mu\nu}R/2=8\pi T_{\mu\nu}\neq0$.

\begin{acknowledgments}
I am greatly indebted to a reviewer for kindly pointing out to me that Eq. (\ref{solplane}) is not a new linearised vacuum solution, but a gauge transformation of the Schwarzschild metric to first order by considering the Regge-Wheeler formalism, and I wish to thank Olivier Sarbach for being very helpful in showing me the explicit gauge transformation which eliminates the perturbation terms in Eq. (\ref{solplane}). Furthermore, I am appreciative of the suggestion by a (presumably different) reviewer that the signature-changing vacuum solutions would contain closed timelike curves.
\end{acknowledgments}

\appendix
\section{Adding non-rotating 2-spheres to slightly deviated plane curve - no higher order vacuum solution}

For second order, the curve:
\begin{eqnarray}\label{hhh}
\tilde{\psi}(r)=\left(0,0,\varepsilon A(r+2R_S)\sqrt{r-R_S}+\varepsilon^2h(r),2\sqrt{R_S(r-R_S)}\right).
\end{eqnarray}
The 3-manifold of revolution:
\begin{eqnarray}
\vec{\sigma}(r,\theta,\phi)&=&\left(r\sin{\theta}\cos{\phi},r\sin{\theta}\sin{\phi},r\cos{\theta}+\varepsilon A(r+2R_S)\sqrt{r-R_S}+\varepsilon^2h(r),\right.\nonumber\\&\ &\left.2\sqrt{R_S(r-R_S)}\right).
\end{eqnarray}
The 3-d spatial metric (to the order of $\varepsilon^2$):
\begin{eqnarray}
ds^2&=&\left(\frac{1}{1-R_S/r}+\frac{3\varepsilon Ar\cos{\theta}}{\sqrt{r-R_S}}+\varepsilon^2\left(\frac{9A^2r^2}{4(r-R_S)}+2h'(r)\cos{\theta}\right)\right)dr^2\nonumber\\&\ &+r^2(d\theta^2+\sin^2{\theta}\ d\phi^2)-\left(\frac{3\varepsilon Ar^2}{\sqrt{r-R_S}}+2\varepsilon^2rh'(r)\right)\sin{\theta}\ drd\theta.
\end{eqnarray}
The (3+1)-d spacetime metric:
\begin{eqnarray}
ds^2&=&-\left((1+\varepsilon C)\left(1-\frac{R_S}{r}\right)+\varepsilon\frac{AR_S}{r}\sqrt{r-R_S}\cos{\theta}+\varepsilon^2\zeta(r,\theta)\right)dt^2\nonumber\\&\ &+\left(\frac{1}{1-R_S/r}+\frac{3\varepsilon Ar\cos{\theta}}{\sqrt{r-R_S}}+\varepsilon^2\left(\frac{9A^2r^2}{4(r-R_S)}+2h'(r)\cos{\theta}\right)\right)dr^2\nonumber\\&\ &+r^2(d\theta^2+\sin^2{\theta}\ d\phi^2)-\left(\frac{3\varepsilon Ar^2}{\sqrt{r-R_S}}+2\varepsilon^2rh'(r)\right)\sin{\theta}\ drd\theta.
\end{eqnarray}
Hence, the non-zero components of the Ricci tensor are (to second order in $\varepsilon$, note that the zeroth and first orders identically vanish, as expected of course):
\begin{eqnarray}
R_{tt}&=&\varepsilon^2\frac{R_S(r-R_S)\cos{\theta}}{2r^5}\left[(2r-R_S)h'-r(r-R_S)h''\right]\nonumber\\&\ &+\varepsilon^2\left[\frac{1}{2r^2}\zeta_{\theta\theta}+\frac{1}{2r^2}\zeta_\theta\cot{\theta}+\frac{(r-R_S)}{2r}\zeta_{rr}+\frac{(4r-5R_S)}{4r^2}\zeta_r+\frac{R_S^2}{4r^3(r-R_S)}\zeta\right]\nonumber\\&\ &+\varepsilon^2\frac{AR_S}{16r^3}\left[4A(6r-7R_S)+18C\sqrt{r-R_S}\cos{\theta}\right.\nonumber\\&\ &\left.\phantom{+\varepsilon^2\frac{AR_S}{16r^3}\ }-\frac{A(72r^2-156R_Sr+85R_S^2)\cos^2{\theta}}{r-R_S}\right]
\end{eqnarray}
\begin{eqnarray}
R_{rr}&=&\varepsilon^2\frac{R_S\cos{\theta}}{2r^3(r-R_S)}\left[(2r-3R_S)h'-3r(r-R_S)h''\right]\nonumber\\&\ &+\varepsilon^2\left[-\frac{r}{2(r-R_S)}\zeta_{rr}+\frac{R_S}{4(r-R_S)^2}\zeta_r-\frac{R_S(4r-3R_S)}{4r(r-R_S)^3}\zeta\right]\nonumber\\&\ &+\varepsilon^2\frac{AR_S}{16r(r-R_S)}\left[-12A(r-R_S)+6C\sqrt{r-R_S}\cos{\theta}\right.\nonumber\\&\ &\left.\phantom{+\varepsilon^2\frac{AR_S}{16r(r-R_S)}}\ +\frac{A(36r^2-68R_Sr+33R_S^2)\cos^2{\theta}}{r-R_S}\right]
\end{eqnarray}
\begin{eqnarray}
R_{\theta\theta}&=&-\varepsilon^2\frac{R_S\cos{\theta}}{r^2}\left[R_S h'+r(r-R_S)h''\right]\nonumber\\&\ &+\varepsilon^2\left[-\frac{r}{2(r-R_S)}\zeta_{\theta\theta}-\frac{r}{2}\zeta_r+\frac{R_S}{2(r-R_S)}\zeta\right]\nonumber\\&\ &+\varepsilon^2\frac{AR_S}{8(r-R_S)}\left[2A(3r-2R_S)-6C\sqrt{r-R_S}\cos{\theta}+A(18r-23R_S)\cos^2{\theta}\right]
\end{eqnarray}
\begin{eqnarray}
\frac{R_{\phi\phi}}{\sin^2{\theta}}&=&-\varepsilon^2\frac{R_S\cos{\theta}}{r^2}\left[R_S h'+r(r-R_S)h''\right]\nonumber\\&\ &+\varepsilon^2\left[-\frac{r}{2(r-R_S)}\zeta_\theta\cot{\theta}-\frac{r}{2}\zeta_r+\frac{R_S}{2(r-R_S)}\zeta\right]\nonumber\\&\ &-\varepsilon^2\frac{3AR_S}{8(r-R_S)}\left[-8A(r-R_S)+2C\sqrt{r-R_S}\cos{\theta}+AR_S\cos^2{\theta}\right]
\end{eqnarray}
\begin{eqnarray}
R_{r\theta}&=&R_{\theta r}=\varepsilon^2\frac{R_S\sin{\theta}}{2r^2}h'+\varepsilon^2\left[\frac{(2r-R_S)}{4(r-R_S)^2}\zeta_\theta-\frac{r}{2(r-R_S)}\zeta_{r\theta}\right]\nonumber\\&\ &+\varepsilon^2\frac{AR_S\sin{\theta}}{8r(r-R_S)^2}\left[6C(r-R_S)^{3/2}-A(18r^2-43R_Sr+24R_S^2)\cos{\theta}\right].
\end{eqnarray}

The same useful relations as in the case of the first order equations can be formed here. Firstly:
\begin{eqnarray}
R_{\theta\theta}-\frac{R_{\phi\phi}}{\sin^2{\theta}}=\varepsilon^2\frac{1}{4(r-R_S)}[2r\zeta_{\theta}\cot{\theta}-2r\zeta_{\theta\theta}-A^2R_S(9r-10R_S)\sin^2{\theta}],
\end{eqnarray}
and $R_{\mu\nu}=0$ implies that
\begin{eqnarray}
\zeta(r,\theta)&=&\zeta_1(r)+\zeta_2(r)\cos{\theta}-\frac{A^2R_S(9r-10R_S)}{4r}\cos^2{\theta}.
\end{eqnarray}
Using $2r(\zeta_{\theta}\cot{\theta}-\zeta_{\theta\theta})=A^2R_S(9r-10R_S)\sin^2{\theta}$, the second useful relation is:
\begin{eqnarray}
r(r-R_S)R_{rr}+\frac{r^3}{(r-R_S)}R_{tt}+2R_{\theta\theta}&=&\varepsilon^2\frac{2R_S\cos{\theta}}{r^2}[(r-2R_S)h'-2r(r-R_S)h'']\nonumber\\&\ &+\varepsilon^2\frac{9}{2}A^2R_S,
\end{eqnarray}
so that $R_{\mu\nu}=0$ gives
\begin{eqnarray}
2r(r-R_S)h''-(r-2R_S)h'=\frac{9}{4}A^2r^2\sec{\theta},
\end{eqnarray}
which has no solution since $h$ is a function of $r$ alone and independent of $\theta$ \footnote{Recall that $h(r)$ is a perturbation of the straight line $\tilde{\psi}(r)$ in Eq. (\ref{hhh}) to cause it to be a plane curve. A curve is intrinsically one-dimensional and only depends on one variable.} unless of course $A=0$ which implies that there is no first order perturbation and the second order perturbation simply then becomes the effective first order perturbation. In fact, looking at the third corresponding useful relation with the use of $\zeta(r,\theta)=\zeta_1(r)+\zeta_2(r)\cos{\theta}-A^2R_S(9r-10R_S)\cos^2{\theta}/4r$:
\begin{eqnarray}
&\ &\frac{r(r-R_S)}{4}R_{rr}+\frac{r^3}{4(r-R_S)}R_{tt}-\frac{1}{2}R_{\theta\theta}-R_{r\theta}(r\cot{\theta})=\nonumber\\&\ &\varepsilon^2\frac{4r(r-R_S)\zeta_1'-4R_S\zeta_1-15A^2R_S(r-R_S)}{8(r-R_S)}\nonumber\\&\ &+\varepsilon^2\frac{[-2(r-R_S)R_Sr\zeta_2'-R_S(r-2R_S)\zeta_2+3A^2R_S(r-R_S)(3r-2R_S)\cos{\theta}]\cos{\theta}}{2(r-R_S)^2}.
\end{eqnarray}
From $R_{\mu\nu}=0$, collecting terms without any $\theta$-dependence gives
\begin{eqnarray}
\zeta_1(r)=E\left(1-\frac{R_S}{r}\right)&+&\frac{15A^2R_S(r-R_S)}{4r}\ln{(r-R_S)}.
\end{eqnarray}
Collecting terms with the $\theta$-dependence, there is no solution to $\zeta_2$ since it is only a function of $r$ and independent of $\theta$ (again, unless $A=0$).

\bibliographystyle{spphys}       
\bibliography{Citation}

\end{document}